\documentclass[pdflatex,sn-mathphys-num,iicol]{sn-jnl}

\usepackage{graphicx}%
\usepackage{multirow}%
\usepackage{amsmath,amssymb,amsfonts}%
\usepackage{amsthm}%
\usepackage{mathrsfs}%
\usepackage[title]{appendix}%
\usepackage{xcolor}%
\usepackage{textcomp}%
\usepackage{manyfoot}%
\usepackage{booktabs}%
\usepackage{algorithm}%
\usepackage{algorithmicx}%
\usepackage{algpseudocode}%
\usepackage{listings}%

\usepackage{physics}
\usepackage{subcaption} 

\theoremstyle{thmstyleone}%
\theoremstyle{thmstyletwo}%

\theoremstyle{thmstylethree}%

\begin{document}

\title[Entropy of the cell fluid model with Curie-Weiss interaction]{Entropy of the cell fluid model with Curie-Weiss interaction}


\author*{\fnm{R.~V.} \sur{Romanik}}\email{romanik@icmp.lviv.ua}

\author{\fnm{O.~A.} \sur{Dobush}}\email{dobush@icmp.lviv.ua}

\author{\fnm{M.~P.} \sur{Kozlovskii}}\email{mpk@icmp.lviv.ua}

\author{\fnm{I.~V.} \sur{Pylyuk}}\email{piv@icmp.lviv.ua}

\author{\fnm{M.~A.} \sur{Shpot}}\email{shpot.mykola@gmail.com}
\affil{\orgname{Yukhnovskii Institute for Condensed Matter Physics of the National Academy of Sciences of Ukraine}, \orgaddress{
		\city{Lviv}, \postcode{79011}, \country{Ukraine}}}


\abstract{Entropy of the cell fluid model with Curie-Weiss interaction is obtained in analytical form as a function of temperature and chemical potential.
A parametric equation is derived representing the entropy as a function of density. Features of both the entropy per particle and the entropy per cell are investigated at the entropy-density and entropy-chemical potential planes. The considered cell model is a multiple-occupancy model and possesses an infinite sequence of first-order phase transitions at sufficiently low temperatures. We find that the entropy exhibits pronounced minima at around integer-valued particle densities, which may be a generic feature of multiple-occupancy models.}




\maketitle

\section{Introduction}

Multiple-occupancy lattice-gas models have been studied much less frequently than classical lattice gases~\cite{LY52,Stanley71}, where each cell can contain at most one particle. Nevertheless, they represent a valuable class of models capable of reproducing the essential thermodynamic features of fluids with soft or penetrable interactions~\cite{Stillinger76,LWL98,FHL04,FL18,LYZ21,WLWLX24}. From the statistical-mechanical point of view, such systems constitute an interesting theoretical object that exhibits nontrivial collective behavior while remaining analytically tractable within mean-field or Curie-Weiss-type approximations~\cite{KKD20,KD22}.

Among the thermodynamic quantities characterizing these systems, entropy plays a central role, as it provides a direct measure of configurational degeneracy and is closely related to the onset of phase transitions. Although some studies~\cite{dOG78,FHL04,Prestipino14,PGT15,FL18,LYZ21,KD22} have addressed the thermodynamic and phase properties of multiple-occupancy models, their entropic characteristics remain largely unexplored. In the present work, we aim to address this gap by deriving explicit analytical expressions for the entropy of the multiple-occupancy cell fluid model with Curie-Weiss-type attraction. Because of its mean-field character, the Curie-Weiss interaction allows for an exact thermodynamic solution, providing a convenient analytical reference for multiple-occupancy systems.

Following Stanley~\cite[Appendix A]{Stanley71}, we interpret the lattice gas as a cell fluid, that is, as a system in which a particle center is located in a cell, and particle positions are not restricted to a discrete set of sites in a lattice. In our case, each cell may contain an arbitrary number of particles, and the  interaction combines both short-range repulsion and long-range mean-field attraction. The present study extends the results of~~\cite{KKD20,KD22,RDKPS25arxiv} by focusing on the entropy and its dependence on chemical potential, particle density, and temperature.

The paper is organized as follows. Section~\ref{sec:model} briefly describes the model. Section~\ref{sec:res} presents exact entropy expressions, and illustrates the results numerically and graphically. Section~\ref{sec:dis} discusses their physical interpretation and qualitative similarity with other systems.

\section{Model and Definitions}
\label{sec:model}
An open system of point particles is considered in volume $V\subset\mathbb R^3$ in three space dimensions. The total volume $V$ is partitioned into $N_v$ non-overlapping congruent cubic cells $\Delta_l$, $l\in\{1,...,N_v\}$, each of volume $v$, such that $V$ is the union of all $\Delta_l$'s:
\begin{eqnarray}\label{volume}
	V = \bigcup_{l=1}^{N_v}\Delta_l,\qquad
	\Delta_l \cap \Delta_{l'} = \emptyset, \text{ if } l \neq l',
	\\ \qquad\mbox{and}\quad
	V = v N_v.
\end{eqnarray}
The two-particle interaction energy is defined as
\begin{equation}
	\label{def:curie-weiss-pot}
	\Phi_{N_v}(\vb{r}_i, \vb{r}_j) = -\frac{J_1}{N_v} + J_2\sum_{l=1}^{N_v} \mathbb{I}_{\Delta_l}(\vb{r}_i) \mathbb{I}_{\Delta_l}(\vb{r}_j).
\end{equation}
The first term in $\Phi_{N_v}$ describes a global Curie-Weiss (mean-field-like) attraction for any pair of particles in the system.
The strength of this attraction is controlled by an energy parameter $J_1 > 0$. The second term in $\Phi_{N_v}$ describes a local repulsion between two particles contained within the same cell $\Delta_l$ and is characterized by the parameter $J_2 > 0.$
Here, $\mathbb{I}_{\Delta_l}(\vb{r})$ is the indicator function of a cell $\Delta_l$,
\begin{equation}
	\label{def:I}
	\mathbb{I}_{\Delta_l} (\vb{r}) = \left\{
	\begin{array}{ll}
		1, \quad \vb{r} \in \Delta_l,
		\\
		0, \quad \vb{r} \notin \Delta_l.
	\end{array}
	\right.
\end{equation}
We refer the reader to Refs.~\cite{KKD20,KD22,RDKPS25arxiv} for a more rigorous definition of the model. The main features of the model are the following:
\begin{itemize}
	\item the model possesses an exact solution in the thermodynamic limit;
	
	\item for sufficiently low temperatures, the model exhibits an infinite sequence of first-order phase transitions between fluid phases of successively increasing density;
	
	\item the phase transition lines between each pair of neighboring phases terminate at respective critical points, and therefore an infinite number of critical points exist in the model.
	
\end{itemize}
The phase diagrams in both the temperature-density and pressure-temperature planes are presented in~\cite{KD22}.

One of the main results of~\cite{RDKPS25arxiv} is the derivation of explicit expressions for the entropy of the model as a function of temperature and chemical potential. In the next Section~\ref{sec:res}, we analyze these expressions for the entropy in detail. 

We use the following standard dimensionless variables: the reduced temperature $T^*=k_{\mathrm{B}}T/J_1$, the reduced pressure $P^* = P v/J_1$, the reduced chemical potential $\mu^* = \mu/J_1$, the reduced particle density $\rho^* = \frac{\langle N \rangle}{V} v = \frac{\langle N \rangle}{N_v}$, with $T$ being the absolute temperature, $k_{\mathrm{B}}$ the Boltzmann constant, $P$ the pressure, $\mu$ the chemical potential, and $\langle N \rangle$ the average number of particles, where the brackets denote a grand-canonical ensemble average. For the entropy $S$, the corresponding reduced quantity is
\begin{equation}
	\label{def:S1}
	S^* = \frac{S}{k_{\mathrm{B}} \langle N \rangle}.
\end{equation}
We call $S^*$ the \textit{entropy per particle}. For lattice models, the \textit{entropy per cell} (or the \textit{entropy per site}) $S^*_v$ is also defined as
\begin{equation}
	\label{def:S2}
	S^*_v = \frac{S}{k_{\mathrm{B}} N_v}.
\end{equation}
The thermodynamic formula for the entropy per particle is
\begin{eqnarray}
	S^* & = & \frac{1}{\rho^*}\left(\frac{\partial P^*}{\partial T^*}\right)_{{\mu^*}}
	= \frac{(\partial P^* / \partial T^*)_{{\mu^*}}}{(\partial P^* / \partial \mu^*)_{T^*}}.
\end{eqnarray}
Using Eqs.~\eqref{def:S1} and \eqref{def:S2}, we obtain the entropy per cell
\begin{equation}
	\label{eq:S2_via_rho_S1}
	S^*_v = \rho^* S^* = \left(\frac{\partial P^*}{\partial T^*}\right)_{\mu^*}.
\end{equation} 

\section{Results}\label{sec:res}
From~\cite{RDKPS25arxiv}, the reduced entropy per particle can be written in the analytical form
\begin{eqnarray}
	\label{eq:entropy1}
	S^*(T^*,\mu^*) = \left(\frac{3}{2} - \frac{\mu^*}{T^*}\right) 
	\nonumber\\
	- \frac{1}{T^*}\frac{K_1}{K_0} + \frac{K_0 \ln K_0}{K_1} + \frac{a}{2T^*} \frac{K_2}{K_1},
\end{eqnarray}
where $a = J_2/J_1$, and the special functions $K_j$, with $j=0, 1, 2, \ldots$, depend on the temperature $T^*$ and chemical potential $\mu^*$:
\begin{eqnarray}
	\label{def:Kj}
	K_j(T^*,\mu^*;\bar{y}_{\mathrm{max}}) =  \sum_{n=0}^{\infty} \frac{n^j (v^* T^{*3/2})^n}{n!}
	\nonumber\\
	\times \exp[\left(\bar{y}_{\mathrm{max}}+\frac{\mu^*}{T^*}\right)n - \frac{a}{2T^*}n^2].
\end{eqnarray}
Here $v^* = v/\lambda^3$, $\lambda = (2\pi\hbar^2/mJ_1)^{1/2}$, with $\hbar$ the Planck constant, $m$ the mass of a particle. A quantity analogous to $v^*$ appears in~\cite{PGT15} in the context of a model that also allows for multiple occupancy.

In \eqref{def:Kj}, the quantity $\bar{y}_{\mathrm{max}} = \bar{y}_{\mathrm{max}}(T^*,\mu^*)$ is a function of $T^*$ and $\mu^*$. It is determined from the extremum condition
\begin{eqnarray}
	\label{cond:max}
	E_1(T^*,\mu^*; \bar{y}_{\mathrm{max}}) & = & 0,
	\nonumber\\
	E_2(T^*,\mu^*;\bar{y}_{\mathrm{max}}) & < & 0,
\end{eqnarray}
with functions $E_1$ and $E_2$ defined as
\begin{equation}
	\label{def:E1}
	E_1(T^*,\mu^*;y) = -T^* y + \frac{K_1(T^*,\mu^*;y)}{K_0(T^*,\mu^*;y)};
\end{equation}
\begin{eqnarray}
	E_2(T^*,\mu^*;y) = -T^*+\frac{K_2(T^*,\mu^*;y)}{K_0(T^*,\mu^*;y)}
	\nonumber\\
	-\left[\frac{K_1(T^*,\mu^*;y)}{K_0(T^*,\mu^*;y)}\right]^2\!.
\end{eqnarray}

When the ideal-gas contribution to the chemical potential, $\mu_{\mathrm{id}} = k_{\mathrm{B}}T \ln(N\Lambda^3/V)$, is taken into account, it becomes evident that the first term in~\eqref{eq:entropy1} 
\begin{equation}
	S^*_{\mathrm{id}} = \frac{3}{2} - \frac{\mu^*}{T^*}
\end{equation}
represents the entropy of a system of noninteracting molecules arranged on a lattice, subject to the strict single-occupancy constraint (one particle per cell), see~\cite[(47.4)]{Hill56}.

The explicit formula for the entropy per cell is
\begin{eqnarray}
	\label{eq:entropy2}
	S^{*}_v(T^*,\mu^*) = \left(\frac{3}{2}	- \frac{\mu^*}{T^*}\right)\frac{K_1}{K_0} 
	\nonumber\\
	- \frac{1}{T^*}\frac{K_1^2}{K_0^2} + \ln K_0 + \frac{a}{2T^*} \frac{K_2}{K_0}.
\end{eqnarray}
Equations~\eqref{eq:entropy1} and~\eqref{eq:entropy2} express the entropy as a function of temperature and chemical potential. To obtain the entropy dependence on the density $\rho^*$, we should account for the following equation \cite[(42)]{RDKPS25arxiv}:
\begin{equation}
	\label{eq:rho_in_y}
	\rho^*=\frac{K_1(T^*,\mu^*;\bar{y}_{\mathrm{max}})}{K_0(T^*,\mu^*;\bar{y}_{\mathrm{max}})}.
\end{equation}

Let us now represent the analytical results for entropy in the form of parametric equations. Noticing that by~\eqref{cond:max} and~\eqref{def:E1}
\begin{equation}
	\bar{y}_{\mathrm{max}} = \frac{1}{T^*}\frac{K_1(T^*,\mu^*;y)}{K_0(T^*,\mu^*;y)}
\end{equation}
and applying the idea taken from~\cite{KD22}, we introduce a new function
\begin{equation}
	\bar{z}_{\mathrm{max}}(T^*,\mu^*) = \bar{y}_{\mathrm{max}} + \frac{\mu^*}{T^*},
\end{equation}
and rewrite the entropy in the form
\begin{equation}
	\label{eq:S_in_z}
	S^*(T^*;\bar{z}_{\mathrm{max}}) = \frac{3}{2} - \bar{z}_{\mathrm{max}} + \frac{\tilde{K}_0 \ln \tilde{K}_0}{\tilde{K}_1} + \frac{a}{2T^*} \frac{\tilde{K}_2}{\tilde{K}_1},
\end{equation}
where the special functions $\tilde{K}_j$ are now written as
\begin{eqnarray}
	\tilde{K}_j(T^*;\bar{z}_{\mathrm{max}}) = \sum_{n=0}^{\infty} \frac{n^j (v^* T^{*3/2})^n}{n!} 
	\nonumber\\
	\times \exp[\bar{z}_{\mathrm{max}}n - \frac{a}{2T^*}n^2].
\end{eqnarray}
In terms of functions $\tilde{K}_j$ and $\bar{z}_{\mathrm{max}}$, the chemical potential is expressed as
\begin{equation}
	\label{eq:mu_in_z}
	\mu^* = T^* \bar{z}_{\mathrm{max}} - \frac{\tilde{K}_1(T^*;\bar{z}_{\mathrm{max}})}{\tilde{K}_0(T^*;\bar{z}_{\mathrm{max}})}.
\end{equation}
At any given temperature $T^*$, Eqs.~\eqref{eq:S_in_z} and~\eqref{eq:mu_in_z} can be formally considered as a parametric equation for the entropy $S^*$ as a function of $\mu^*$, with $\bar{z}_{\mathrm{max}}$ being the parameter.

Similarly, if Eq.~\eqref{eq:rho_in_y} is rewritten as
\begin{equation}
	\label{eq:rho_in_z}
	\rho^*(T^*;\bar{z}_{\mathrm{max}}) = \frac{\tilde{K}_1(T^*;\bar{z}_{\mathrm{max}})}{\tilde{K}_0(T^*;\bar{z}_{\mathrm{max}})},
\end{equation}
then Eqs.~\eqref{eq:S_in_z} and~\eqref{eq:rho_in_z} constitute a parametric equation for the entropy $S^*$ as a function of $\rho^*$, at any given $T^*$ and with $\bar{z}_{\mathrm{max}}$ being the parameter.

The explicit expression for the entropy per cell in terms of $\bar{z}_{\mathrm{max}}$ and $\tilde{K}_j$ reads
\begin{equation}
	\label{eq:S2_in_z}
	S^*_v(T^*;\bar{z}_{\mathrm{max}}) = \left(\frac{3}{2} - \bar{z}_{\mathrm{max}}\right) \frac{\tilde{K}_1}{\tilde{K}_0} + \ln \tilde{K}_0 + \frac{a}{2T^*} \frac{\tilde{K}_2}{\tilde{K}_0}.
\end{equation}
Therefore, Eq.~\eqref{eq:S2_in_z} together with~\eqref{eq:rho_in_z} provides the corresponding parametric representation of the entropy $S^*_v$ as a function of $\rho^*$ for a given $T^*$, with $\bar{z}_{\mathrm{max}}$ being the parameter.

\begin{table*}
	\caption{Critical point coordinates and entropies for the first four critical points of the cell fluid model with Curie-Weiss interaction. The numerical calculations were performed for $a=1.2$ and $v^*=5.0$.}
	\label{tab1}
	\centering
	\begin{tabular}{ccccccc}
		$n$ & $T^*_c$ & $\rho^*_c$ & $\mu^*_c$ & $S^*_c$ & $S^*_{v, c}$ & $\bar{z}_c$ \\
		\midrule
		
		1 & 0.254567 & 0.513896 & 0.209380 & 2.43173 & 1.24966 & 2.84119 \\
		\midrule
		2 & 0.261881 & 1.50557 & 0.585097 & 1.34933 & 2.03151 & 7.98327 \\
		
		\midrule
		3 & 0.265254 & 2.50303 & 0.891843 & 0.914901 & 2.29002 & 12.7985 \\
		
		\midrule
		4 & 0.267242 & 3.50191 & 1.16893 & 0.631157 & 2.21025 & 17.4779 \\
		
		\bottomrule
	\end{tabular}		
\end{table*}

Before proceeding with analysis of the entropy behavior, let us recall that the considered model possesses an infinite number of critical points, and thus, an infinite number of first-order phase transitions at sufficiently low temperatures~\cite{KD22}. We call the phase with the lowest density Phase I. Then, the first first-order phase transition leads to a phase with a higher density, Phase II. The second first-order phase transition results in Phase III, and so on to Phase IV, Phase V, etc. The phase diagram in the temperature-density plane is shown in~\cite[Fig.~10]{KD22}, while detailed numerical values of physical quantities at the first five critical points are summarized in \cite[Table~1]{RDKPS25arxiv}. For reference, Table~\ref{tab1} lists the critical values of relevant physical quantities for the first four critical points, and Table~\ref{tab2} presents their values on coexistence lines at temperatures $T^*=0.25$ and $T^*=0.20$. 

\begin{table*}[h]
	\caption{Coexistence-line values for physical quantities. The numerical calculations were performed for $a=1.2$ and $v^*=5.0$.}
	\label{tab2}
	\begin{tabular}{c|c|ccccccc}
		\toprule
		Coexisting phases	&$T^*$	& $\mu^*_{I-II}$	& $\rho^*_I$   & $\rho^*_{II}$ & $S^*_{I}$ & $S^*_{II}$ & $\bar{z}_{I}$ & $\bar{z}_{II}$ \\
		\midrule
		\multirow{2}*{I - II} &
		0.25 & 0.214259 & 0.397996 & 0.628565 & 2.73453 & 2.11856 & 2.44902 & 3.37130 \\
		&0.20 & 0.259145 & 0.142070 & 0.871833 & 3.57349 & 1.18911 & 2.00607 & 5.65489 \\
		
		\midrule
		& & $\mu^*_{II-III}$	& $\rho^*_{II}$   & $\rho^*_{III}$ & $S^*_{II}$ & $S^*_{III}$ & $\bar{z}_{II}$ & $\bar{z}_{III}$ \\
		\midrule
		\multirow{2}*{II - III}
		& 0.25 & 0.589630 & 1.32372 & 1.68613 & 1.36451 & 1.14578 & 7.65338 & 9.10305 \\
		& 0.20 & 0.598952 & 1.12438 & 1.88058 & 0.996714 & 0.601031 & 8.61664 & 12.3977 \\
		
		\midrule
		& & $\mu^*_{III-IV}$	& $\rho^*_{III}$   & $\rho^*_{IV}$ & $S^*_{III}$ & $S^*_{IV}$ & $\bar{z}_{III}$ & $\bar{z}_{IV}$ \\
		\midrule
		\multirow{2}*{III - IV}
		& 0.25 & 0.891559 & 2.29983 & 2.70532 & 0.874317 &0.733783 & 12.7655 & 14.3875 \\
		& 0.20 & 0.880351 & 2.11796 & 2.88459 & 0.510408 & 0.269009 & 14.9915 & 18.8247 \\
		
		\midrule
		& & $\mu^*_{IV-V}$	& $\rho^*_{IV}$   & $\rho^*_{V}$ & $S^*_{IV}$ & $S^*_{V}$ & $\bar{z}_{IV}$ & $\bar{z}_{V}$\\
		\midrule
		\multirow{2}*{IV - V}
		& 0.25 & 1.16371 & 3.28767 & 3.71550 & 0.565479 & 0.459722 & 17.8055 & 19.5168 \\
		& 0.20 & 1.13801 & 3.11462 & 3.88693 & 0.208441 & 0.0303822 & 21.2632 & 25.1247 \\
		
		\bottomrule
	\end{tabular}
\end{table*}

\begin{figure*}[h]
		\centering
		\begin{subfigure}[b]{0.45\textwidth}
			{\includegraphics[width=\textwidth]{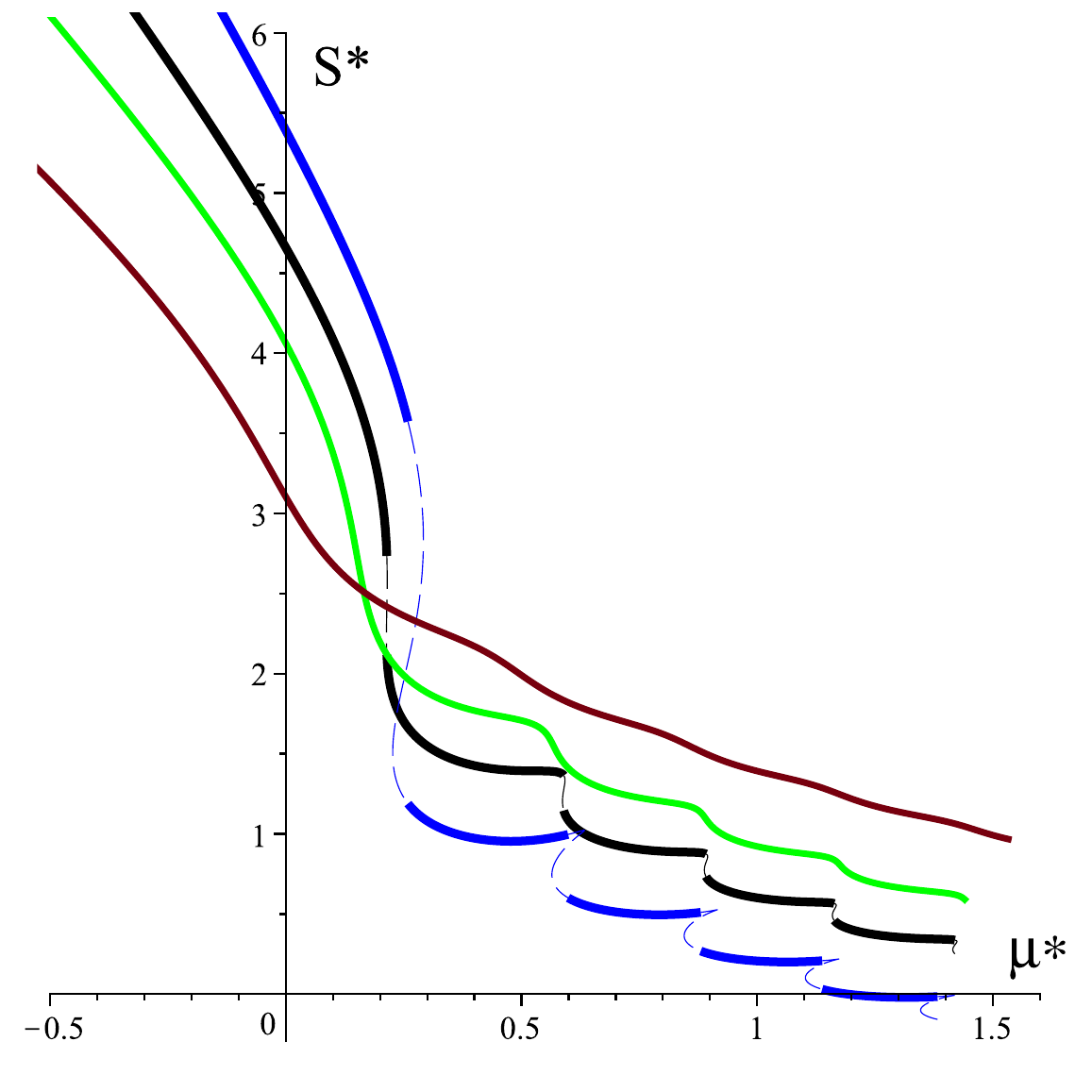}}
			\caption{}
			\label{fig:entropy_a}
		\end{subfigure}
		\hfill
		\begin{subfigure}[b]{0.45\textwidth}
			{\includegraphics[width=\textwidth]{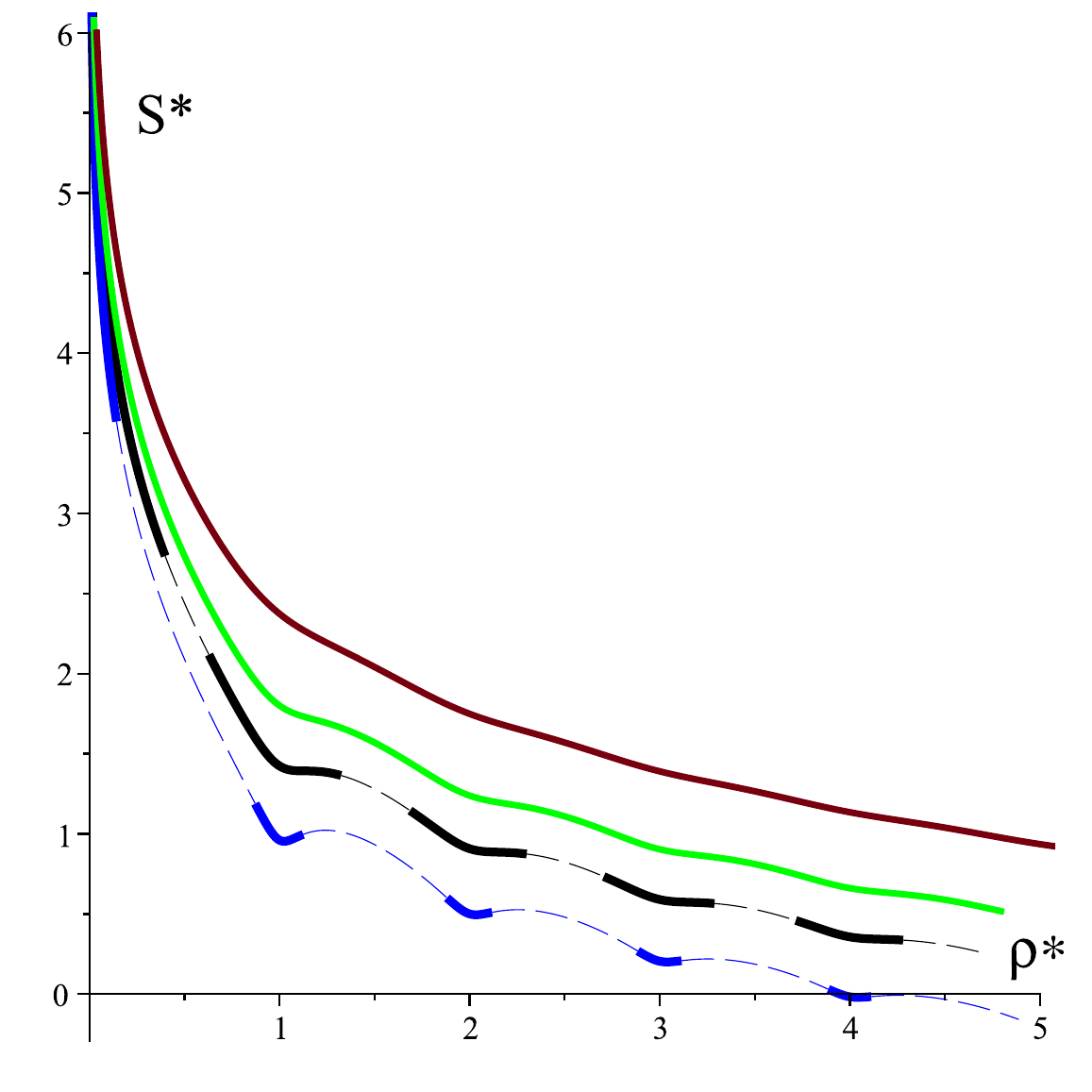}}
			\caption{}
			\label{fig:entropy_b}
		\end{subfigure}
		\begin{subfigure}[b]{0.45\textwidth}
			{\includegraphics[width=\textwidth]{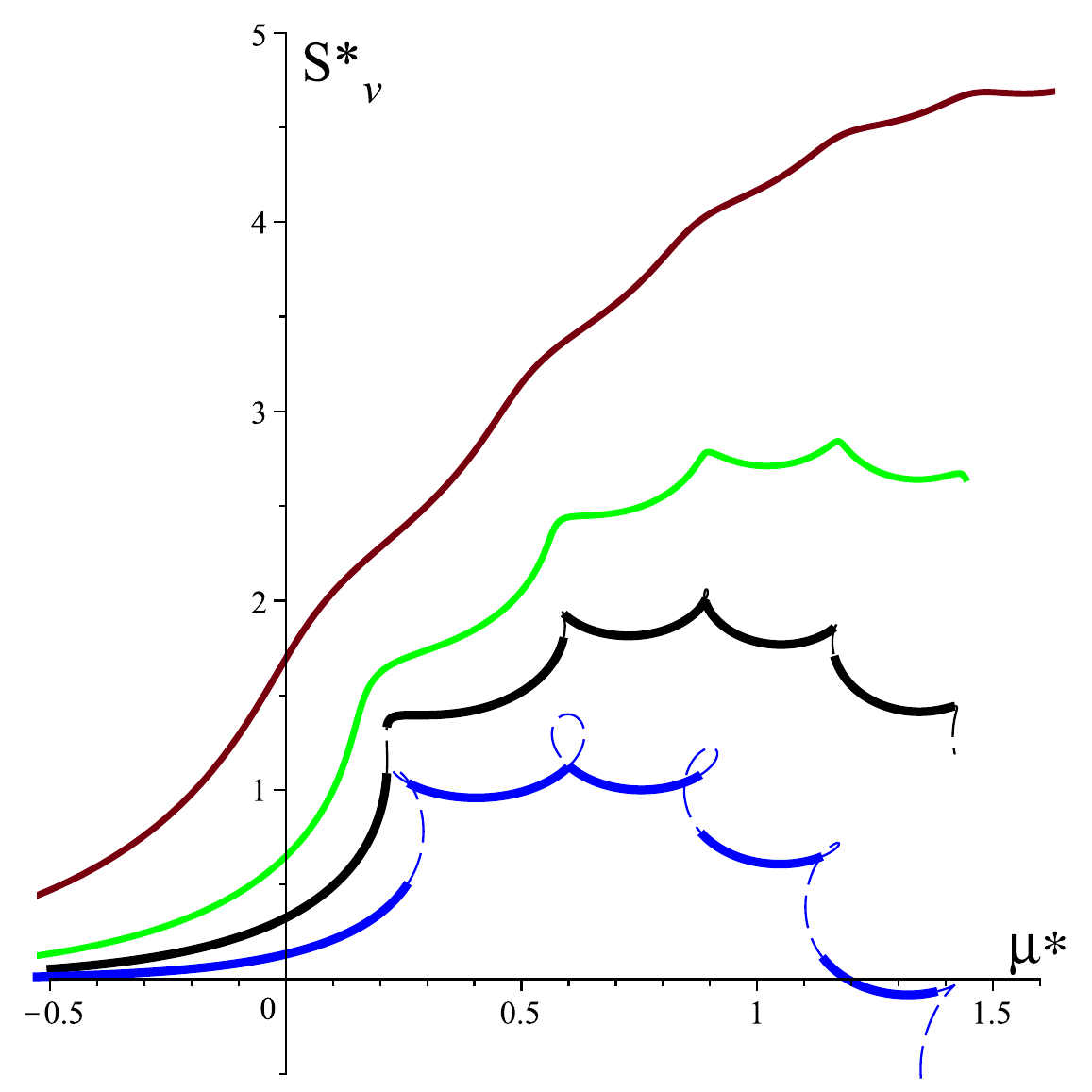}}
			\caption{}
			\label{fig:entropy_c}
		\end{subfigure}
		\hfill
		\begin{subfigure}[b]{0.45\textwidth}
			{\includegraphics[width=\textwidth]{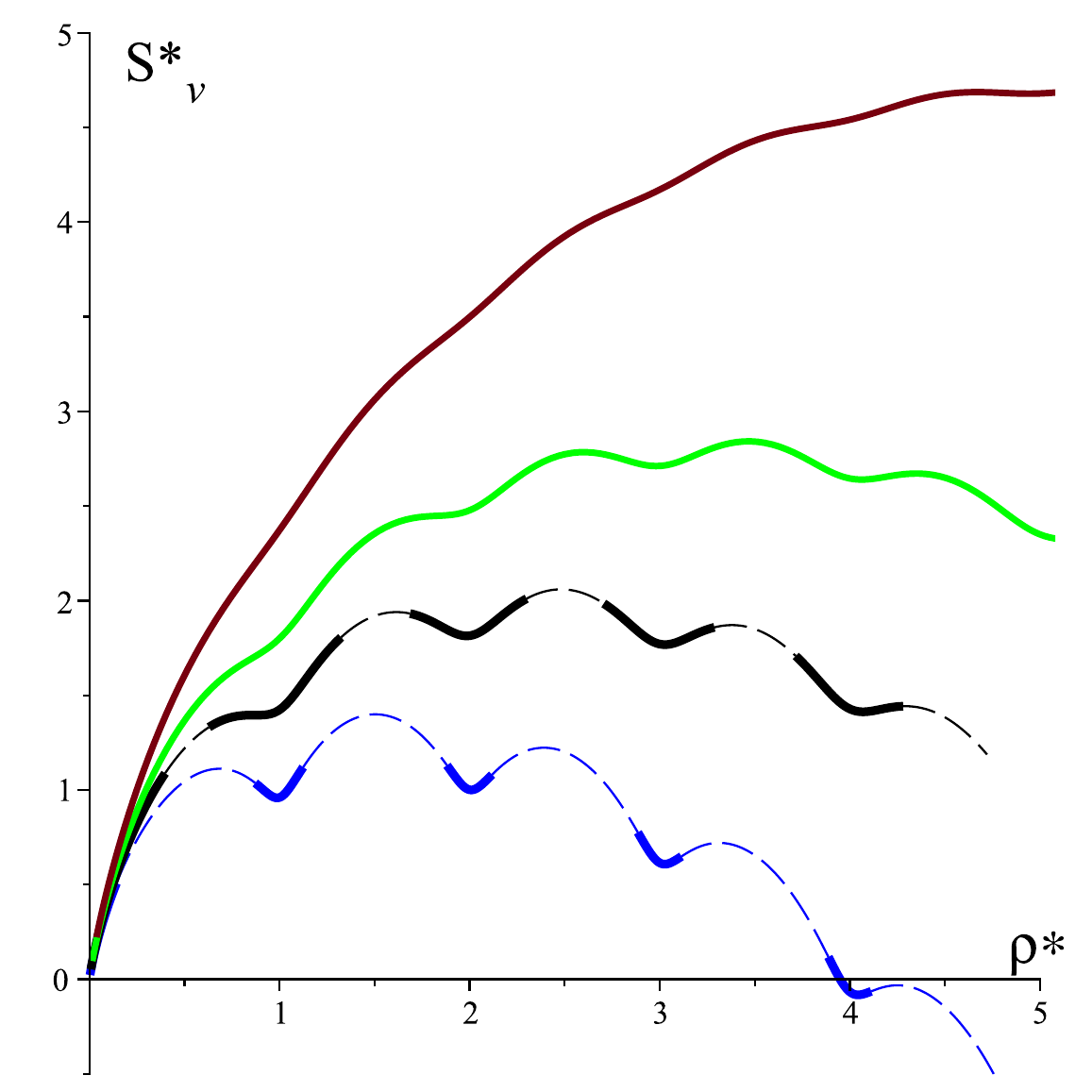}}
			\caption{}
			\label{fig:entropy_d}
		\end{subfigure}
	\caption{Entropy: (\textbf{a}) Entropy per particle versus chemical potential. (\textbf{b}) Entropy per particle versus density. (\textbf{c}) Entropy per cell versus chemical potential. (\textbf{d}) Entropy per cell versus density. In all figures, color curves correspond to the following temperatures: Red - $T^*=0.40$; Green - $T^*=0.30$ Black - $T^*=0.25$; Blue - $T^* = 0.20$. Parameters are taken as $a=1.2$ and $v^*=5.0$.\label{fig:entropy}}
\end{figure*}

The analytical results for the entropy are visualized in Figure~\ref{fig:entropy}. Panels (a) and (b) illustrate the entropy per particle, while panels (c) and (d) correspond to the entropy per cell.
The numerical values for the parameters are taken as $a=1.2$, and $v^*=5.0$. Each panel in Figure~\ref{fig:entropy} shows four temperature curves: the red curve is $T^* = 0.4$, which is much higher than the critical temperatures; the green curve is $T^* = 0.3$, which is slightly higher than the critical temperatures; the black curve is $T^*=0.25$, which is just slightly lower than the critical temperatures; and the blue curve is $T^*=0.20$, which is significantly lower than the critical temperatures. The solid parts of the curves represent the results for the stable phases, while the dashed parts correspond to two-phase regions, i.e. to solutions $\bar{y}$ to $E_1=0$ that may violate the condition of global maxima $E_2 < 0$, see~\eqref{cond:max}. 

The dependence of the entropy per particle on the chemical potential is presented in Figure~\ref{fig:entropy_a}. For the supercritical region, red and green curves, the entropy $S^*$ has a monotonic decreasing behavior with the chemical potential $\mu^*$. For temperatures below the critical ones, we observe that phase II has a lower entropy than Phase I. Consequently, Phase III has a lower entropy than Phase II, and so on. However, within a particular phase, except Phase I, the dependence of the entropy per particle on the chemical potential may not be monotonic. At the phase transitions, the entropy exhibits jumps from a higher value in the phase with a lower chemical potential, to a lower value in the phase with a higher chemical potential.

The dependence of the entropy per particle on the density is illustrated in Figure~\ref{fig:entropy_b}. For supercritical temperatures, the entropy per particle is a decreasing function of the density $\rho^*$. At temperatures lower than the critical ones, the entropy is still a decreasing function of $\rho^*$ in Phase I, but in subsequent phases it exhibits minima around integer values of $\rho^*$. From the plots it is also seen that at the first-order phase transitions, the discontinuities in both entropy and density occur.

We should note in Figure~\ref{fig:entropy_b} that in phases with higher densities, i.e. Phase V and higher, the entropy becomes negative. On the one hand, it is an artifact of the classical treatment of the model. On the other hand, this behavior can also be attributed to quite small values of $v^*$, which resembles the observation by Stillinger~\cite{Stillinger76} for the Gaussian core model, who noted that {\it{``$\ldots$ when $T>0$, the requirement that particles remain confined to a regular lattice as the spacing in that lattice goes to zero creates a diverging negative entropy that could only be offset by an unbounded energy in the same limit.''}}
If the value for $v^*$ is increased, we would observe many more phases with $S^*>0$, before the entropy becomes negative, but nevertheless, for each selection of $v^*$, a density will be reached where the entropy becomes negative.

The dependence of the entropy per cell on the chemical potential is illustrated in Figure~\ref{fig:entropy_c}. At high temperatures, i.e. $T^*=0.4$, the entropy per cell increases as a function of $\mu^*$ for negative and small positive values, reaches a maximum, and then starts decreasing. For temperatures just slightly above the critical ones, i.e. $T^*=0.3$, the behavior of $S^*_v$ is very nontrivial, with noticeable bends around the approaching critical points. At lower temperatures, the entropy per cell increases with $\mu^*$ in Phase I. In Phase II and Phase III, it has a non-monotonic behavior, but generally, these phases possess the highest entropy per cell compared to all other ones. Then, for the consecutive phases, the entropy per cell jumps to lower values at each phase transition, while still having non-monotonic behavior within a single phase. 

The dependence of the entropy per cell on the density is presented in Figure~\ref{fig:entropy_d}. For high temperatures, i.e. $T^*=0.4$, the entropy per cell increases as a function of $\rho^*$ for small values, reaches a maximum, and then starts decreasing. For temperatures just slightly above the critical ones, i.e. $T^*=0.3$, the behavior of $S^*_v$ starts to resemble that below $T_c$, signaling the approach to critical points. At lower temperatures, the entropy per cell increases with $\rho^*$ in Phase I. Then, within each consecutive phase, it exhibits a non-monotonic behavior, and possesses minima at around integer-valued $\rho^*$.
Based on the available data, even though very scarce, we assume that such behavior of the entropy versus density is a generic characteristic of cell (lattice) models allowing for multiple occupancy, see the discussion in the next Section~\ref{sec:dis}.

\section{Discussion}\label{sec:dis}
Published results on entropy in multiple-occupancy lattice models are scarce. To our knowledge, no analogous results exist for classical systems, while several studies have addressed the quantum cases. Interestingly, these quantum models exhibit features remarkably similar to those found in our classical cell fluid model, suggesting a common physical origin -- the allowance for multiple occupancy.

Reference~\cite{dLBKGS11} thoroughly examines both entropy per site and the entropy per cell for the three-dimensional fermionic Hubbard model by the dynamical mean-field theory. In this model, the maximum occupancy (or filling) is 2. Note that their concept of the entropy per site is analogous to our entropy per cell.
The entropy per site as a function of chemical potential is presented in~\cite[Figs.~5 and~6]{dLBKGS11}. The dependence of the entropy per site on density is shown in \cite[Fig.~7]{dLBKGS11}.
That of the entropy per particle on density is illustrated in \cite[Fig.~8]{dLBKGS11}. All figures contain curves for multiple temperatures, and different Hamiltonian parameters of the considered Hubbard model.
However, the dependence of the entropy per particle on chemical potential is not shown. 

Qualitative similarities between our results and the results of~\cite{dLBKGS11} are as follows. The evolution of the entropy per cell for the cell model as a function of $\mu^*$ at high temperature, i.e. $T^*=0.4$ in Figure~\ref{fig:entropy_c}, is similar to the entropy per site versus chemical potential for the Hubbard model at the weak interaction strength, cf. Fig.~5 from~\cite{dLBKGS11}. At the lower temperature, our result at $T^*=0.3$ resembles that of~\cite[Fig.~6]{dLBKGS11} for the Hubbard model at the strong interaction strength. The similarity is particularly evident in the interval $0.7 \leq \mu^* \leq 1.3$ (Figure~\ref{fig:entropy_c}, green curve), where two maxima of $S^*_v$ are observed. The behavior of the entropy per cell versus density, Figure~\ref{fig:entropy_d}, at high temperature, i.e. $T^*=0.4$, is qualitatively the same as the entropy per site versus density of the Hubbard model at the weak interaction strength~\cite[Fig.~7]{dLBKGS11}. At low temperature, the qualitative similarity is observed in the region at $\rho \approx 1.0$, where the dips in the entropy per cell are developed in both models. Finally, we compare the entropy per particle $S^*$, Figure~\ref{fig:entropy_b}, with that of the Hubbard model from~\cite[Fig.~7]{dLBKGS11}. The high temperature regime in Figure~\ref{fig:entropy_b} clearly resembles the weak interaction strength regime in~\cite[Fig.~7]{dLBKGS11}. In turn, in the low-temperature regime, $T^*=0.2$, for Phase II, the entropy per particle develops a minimum at $\rho^* \approx 1$ analogously as in the Hubbard model at the strong interaction strength~\cite[Fig.~7]{dLBKGS11}.

Results similar to those in Ref.~\cite{dLBKGS11} were also reported in~\cite{Campo15,PKF20} for the one-dimensional repulsive Hubbard model, studied using different methods. In both studies, see~\cite[Fig.~4]{Campo15} and~\cite[Fig.21a]{PKF20}, the entropy per site exhibits a single maximum at high temperature, and a pronounced minimum at density $\rho^*=1.0$ and low temperature, which is consistent with our results in Figure~\ref{fig:entropy}.

In~\cite{PKvHT08}, the entropy in the single-band Bose-Hubbard model is studied in one and two dimensions by the quantum Monte Carlo methods. The entropy per site as a function of the filling is presented in \cite[Fig.~2]{PKvHT08} for the homogeneous 1D Bose-Hubbard system. There, the plot is extended up to the filling $n = 3.0$, and at the moment, this is the only study we found that reports the dependence of the entropy per site on the average occupancy number for the density exceeding $\rho^* \approx 2.0$. In~\cite[Fig.~2]{PKvHT08}, we observe two dips at fillings $n = 1.0$ and $n = 2.0$, see the red curve for $U/t = 12$. Analogous dips are observed in our model as well, around $\rho^* \approx 1.0$ and $\rho^* \approx 2.0$, respectively, at $T^*=0.2$, see~Figure~\ref{fig:entropy_d}.

In summary, despite their different physical nature, the allowance for maximum occupancy to be higher than unity leads to remarkably similar features in the mentioned models. The Hubbard model, allowing for $\rho^*_{\mathrm{max}} = 2.0$, develops a minimum in the entropy per site at $\rho^* = 1.0$. The Bose-Hubbard model data, reported up to $\rho^* \leq 3.0$ show the minima for the entropy per site at $\rho^*=1.0$ and at $\rho^* = 2.0$. The cell model with the Curie-Weiss interaction considered in the current work does not have any restriction on the occupancy number, and thus develops the minima in the entropy per cell at $\rho^* \approx 1.0$, $\rho^* \approx 2.0$, $\rho^* \approx 3.0$, and so on. We expect this feature to appear in other multiple-occupancy lattice gas models as well. In particular, the entropy per cell in the double-occupancy lattice gas~\cite{LYZ21,WLWLX24} should have a dip at $\rho^* = 1.0$. The generalized exponential model of index $4$ considered in \cite{Prestipino14,PGT15} in the context of cluster crystals and the lattice gas model of multiple layer adsorption~\cite{dOG78} should both have minima in the entropy per site at integer-valued densities as both models exhibit multiple first-order phase transitions at sufficiently low temperatures, similar to the cell model studied by us. The same arguments apply to the entropy per particle as well.

Additionally, in Appendix~\ref{app:vdW} we present the entropy of the van der Waals fluid in a form analogous to Figure~\ref{fig:entropy}. Note, that a qualitative comparison between the cell model and the van der Waals fluid can be made only for the transition between Phase I and Phase II. This suggests that the first-order phase transition between Phase I and Phase II is of gas-liquid type.

\section{Conclusions}
\label{sec:conclusions}
We have analytically derived explicit expressions for the entropy of a cell fluid model with Curie-Weiss-type interaction and possible multiple occupancy of cells. Within this framework, both the entropy per particle and the entropy per cell were obtained as functions of temperature and chemical potential, and their dependences on density were represented in a parametric form. The analysis demonstrates that the entropy exhibits a sequence of discontinuities at first-order phase transitions, in full correspondence with the infinite cascade of coexistence regions between fluid phases of increasing density.

At temperatures below the critical points, the entropy per particle decreases stepwise with increasing chemical potential, while the entropy per cell exhibits pronounced nonmonotonic behavior within individual phases. In the entropy-density representation, both quantities display distinct minima near integer-valued densities. These minima are directly related to the discrete character of cell occupancy and appear to be a generic thermodynamic signature of multiple-occupancy lattice systems.

The study further reveals that, at sufficiently high densities or small values of the reduced cell volume $v^*$, the entropy may become negative, which is an artifact of the classical treatment analogous to that observed in the Gaussian-core model. Increasing $v^*$ shifts this behavior to higher densities while preserving the overall qualitative pattern.

A qualitative comparison with lattice models of different microscopic origin, such as the fermionic Hubbard and Bose-Hubbard models, shows that all systems allowing for more than single occupancy develop entropy minima at integer densities. The present Curie-Weiss cell model, unrestricted in occupancy number, extends this tendency to an infinite sequence of minima corresponding to each successive phase. This feature suggests a unifying thermodynamic behavior among multiple-occupancy lattice gases (cell fluids) and related soft-matter systems with cluster-forming interactions.


\bmhead{Acknowledgements} This work was supported by the National Research Foundation of Ukraine under the project No. 2023.03/0201.

\bmhead{Data Availability Statement} Data sharing is not applicable to this article as no data sets were generated or analyzed during the current study.

\bmhead{Conflict of Interest Statement} The authors have no conflicts of interest.

\begin{appendices}

\section{Entropy of the van der Waals fluid}
\label{app:vdW}
\begin{figure*}[h]
	\centering
	\begin{subfigure}[b]{0.4\textwidth}
		{\includegraphics[width=\textwidth]{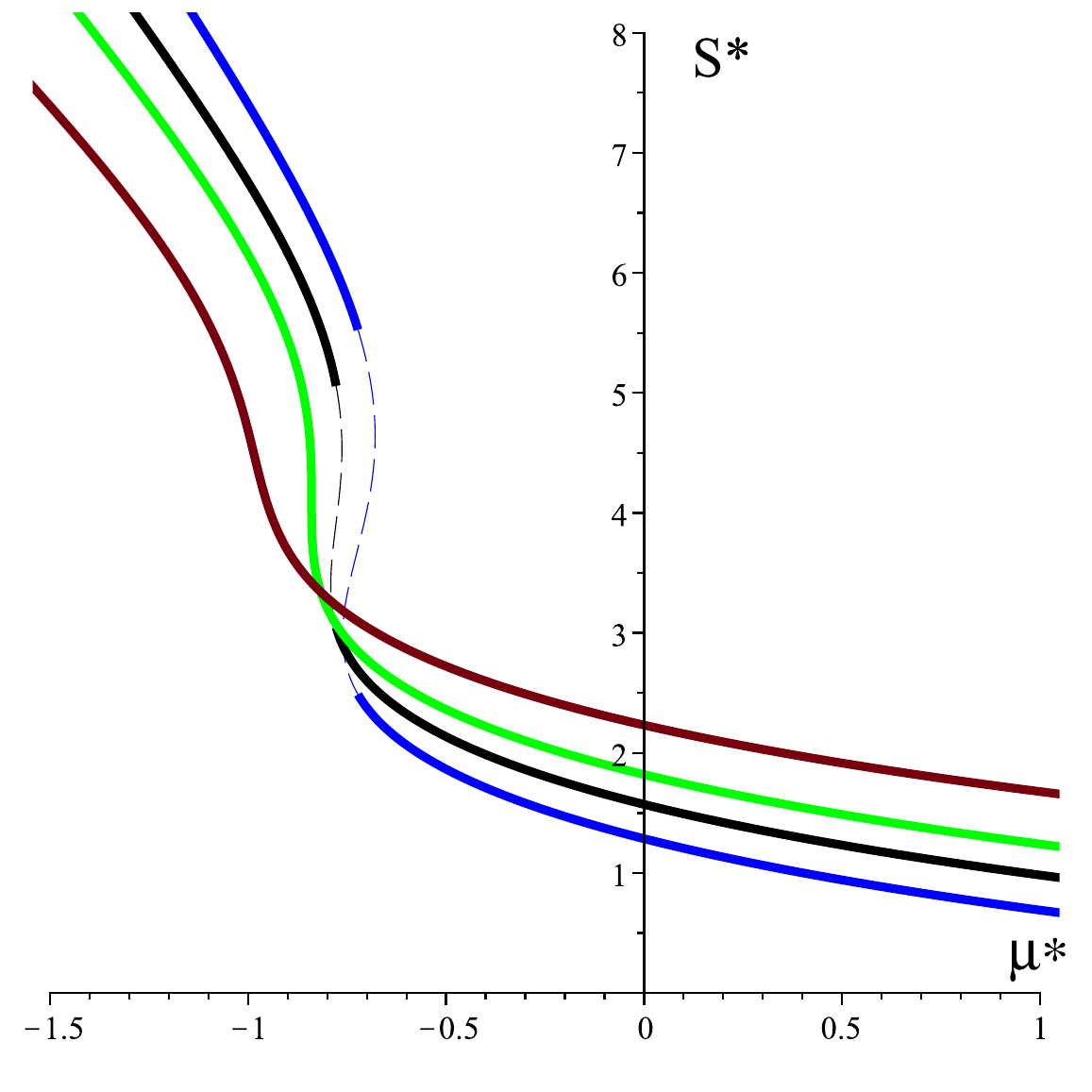}}
		\caption{}
		\label{fig:entropy_vdW_a}
	\end{subfigure}
	\begin{subfigure}[b]{0.4\textwidth}
		{\includegraphics[width=\textwidth]{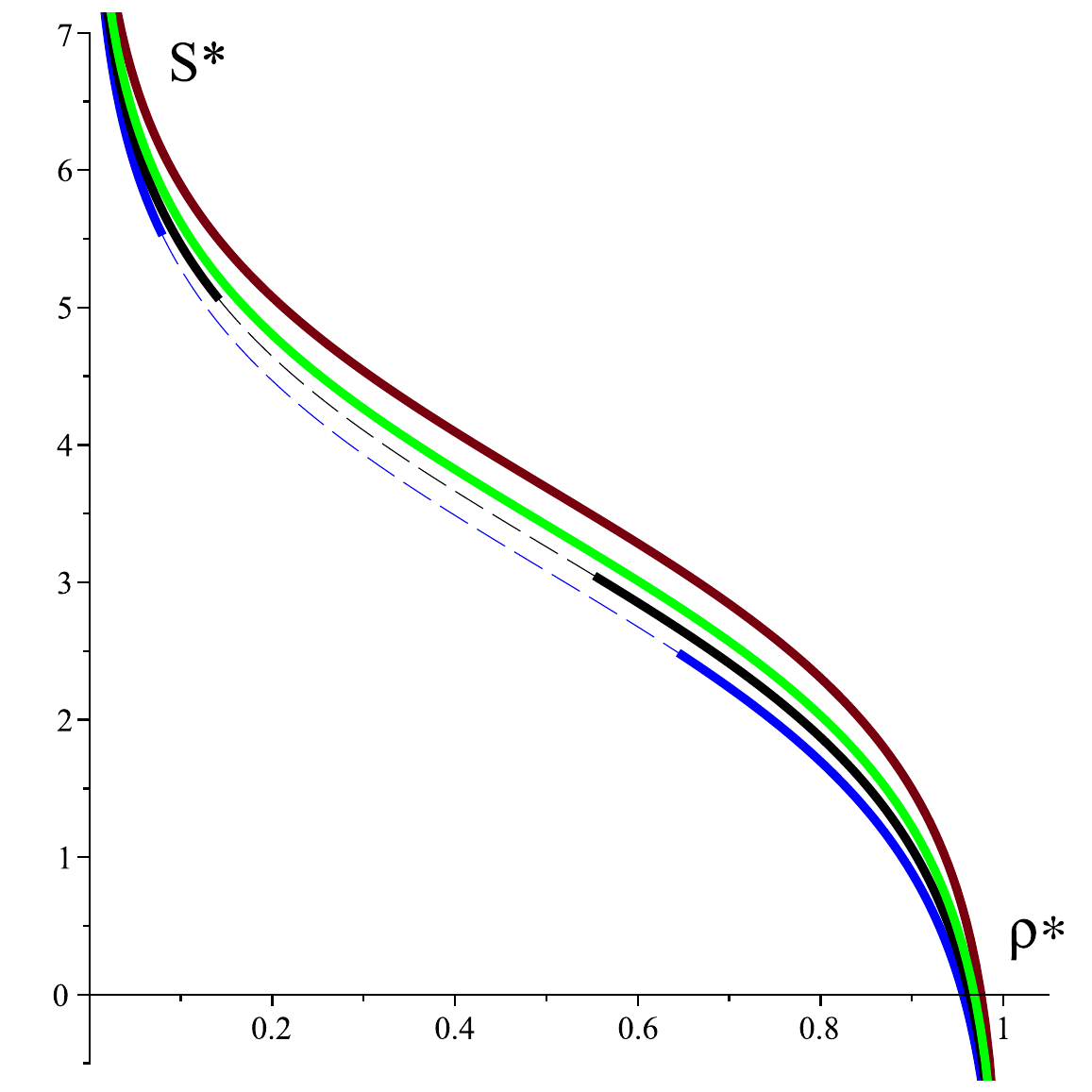}}
		\caption{}
		\label{fig:entropy_vdW_b}
	\end{subfigure}
	\begin{subfigure}[b]{0.4\textwidth}
		{\includegraphics[width=\textwidth]{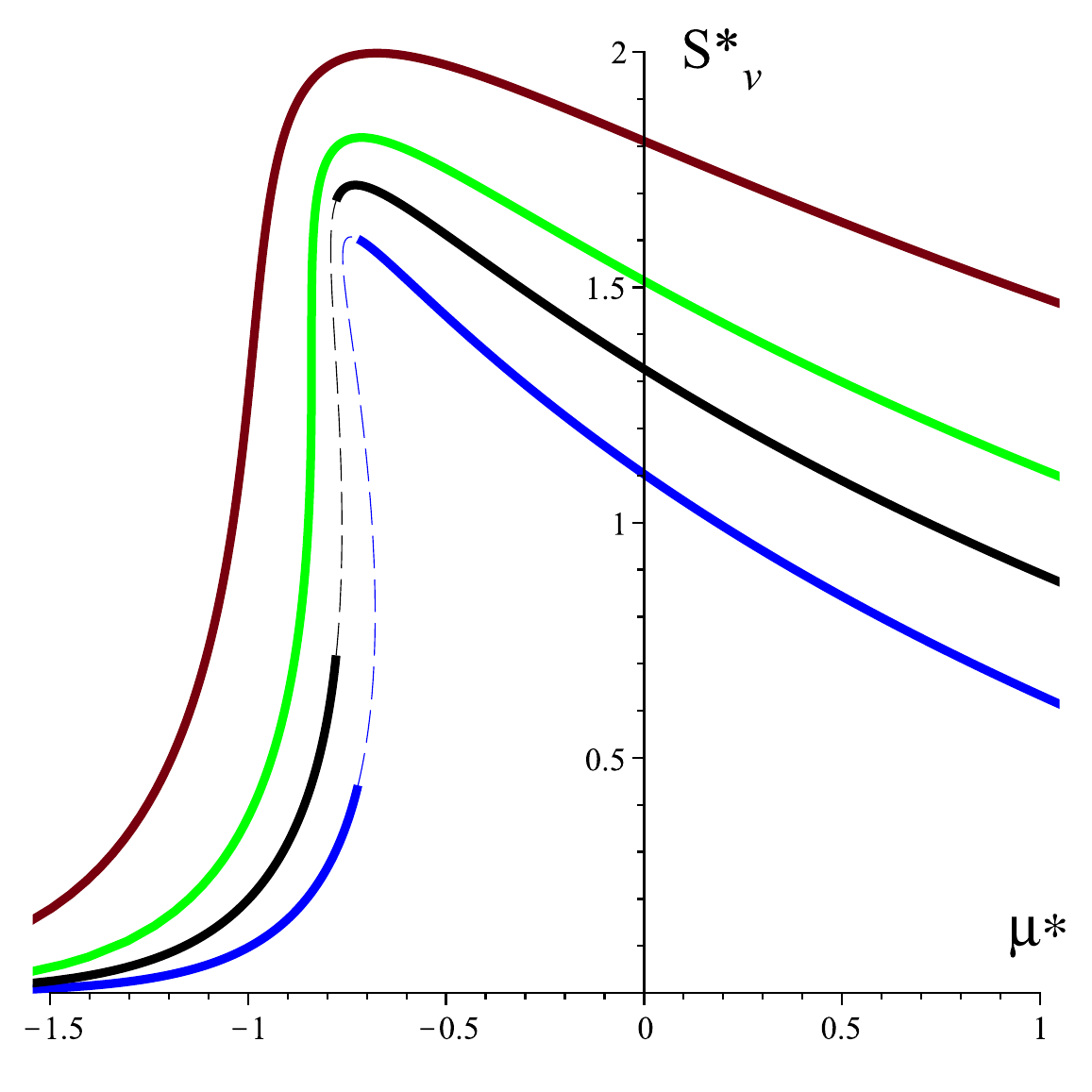}}
		\caption{}
		\label{fig:entropy_vdW_c}
	\end{subfigure}
	\begin{subfigure}[b]{0.4\textwidth}
		{\includegraphics[width=\textwidth]{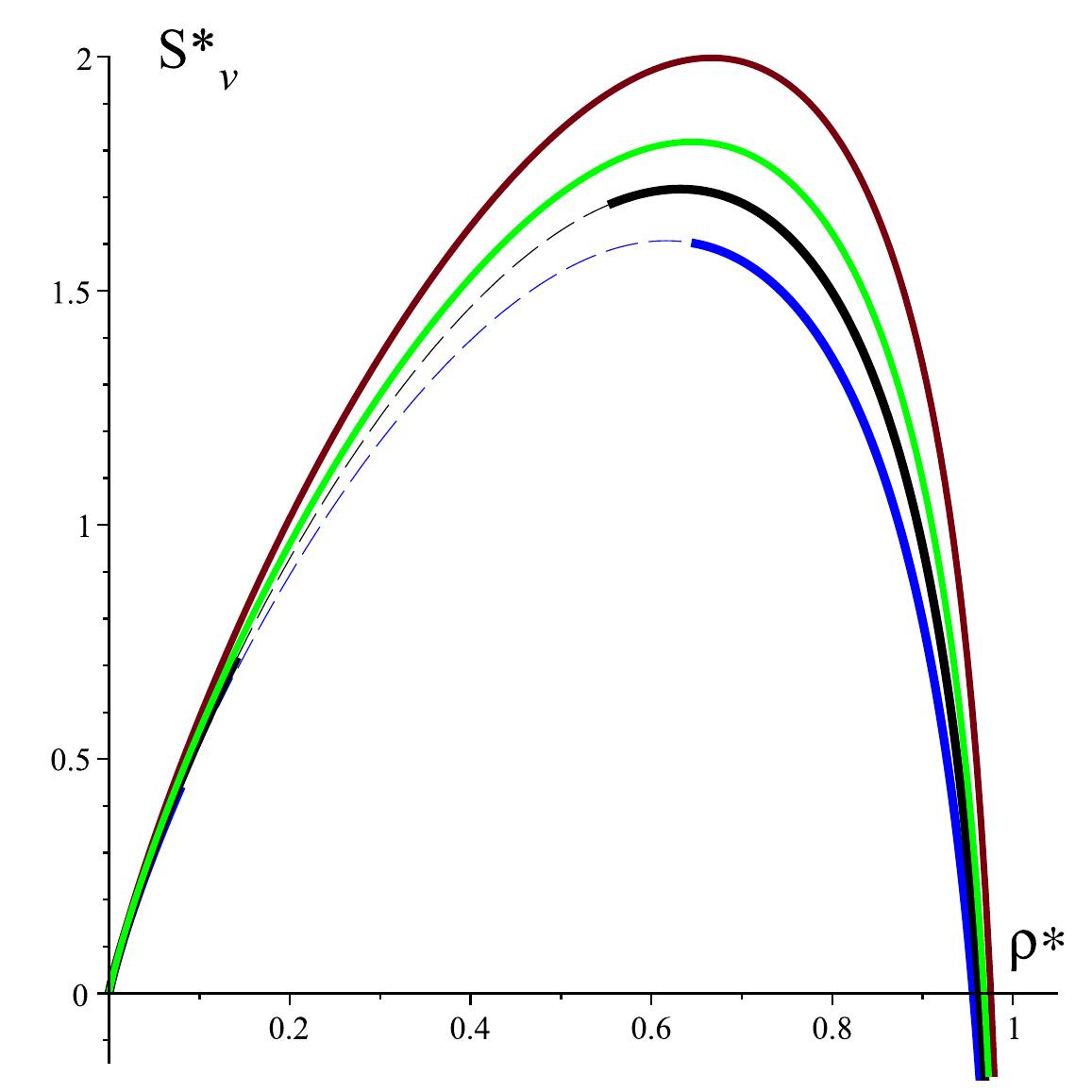}}
		\caption{}
		\label{fig:entropy_vdW_d}
	\end{subfigure}
	\caption{Entropy of the van der Waals fluid: (\textbf{a}) Entropy per particle versus chemical potential. (\textbf{b}) Entropy per particle versus density. (\textbf{c}) Entropy per cell versus chemical potential. (\textbf{d}) Entropy per cell versus density. In all figures, color curves correspond to the following temperatures: Red - $T^*_{\mathrm{vdW}}=0.30$; Green - $T^*_{\mathrm{vdW}}=0.25$ Black - $T^*_{\mathrm{vdW}}=0.225$; Blue - $T^*_{\mathrm{vdW}} = 0.20$. The value of the parameter: $v^*_{\mathrm{vdW}}=5.0$.\label{fig:entropy_vdW}}
\end{figure*} 

The entropy of the van der Waals (vdW) fluid is expressed via density and temperature as~\cite[(55)]{Johnston14}
\begin{equation}
	\label{vdw:S}
	\frac{S}{N k_{\mathrm{B}}} = \ln \left[x_c\hat{\tau}^{3/2}(3-\hat{n})/\hat{n}\right] + \frac{5}{2},
\end{equation}
where $\hat{\tau} = T/T_c$, $\hat{n} = \rho/\rho_c$, and 
\begin{equation}
	x_c \equiv \frac{k_{\mathrm{B}} T_c}{8P_c \Lambda^3_c} = \frac{b}{\Lambda^3_c},
\end{equation}
where $\Lambda_c$ is the de Broglie wavelength at the critical temperature, and $b$ is the ``excluded volume'' parameter of the van der Waals theory. The meaning of $b/\Lambda^3_c$ is similar to the meaning of quantity $v^*$ in our cell theory.

The chemical potential of the vdW fluid as a function of temperature and density is given by~\cite[(70c)]{Johnston14}
\begin{eqnarray}
	\label{vdw:mu}
	\frac{\mu}{k_{\mathrm{B}}T_c} = -\hat{\tau} \ln \left(\frac{3-\hat{n}}{\hat{n}}\right) + \frac{\hat{\tau}\hat{n}}{3-\hat{n}} - \frac{9\hat{n}}{4} 
	\nonumber\\
	- \hat{\tau}\ln(\hat{\tau}^{3/2}) - \hat{\tau}\ln(x_c).
\end{eqnarray}
Eqs.~\eqref{vdw:mu} and~\eqref{vdw:S} can be considered as a parametric equation for the entropy as a function of the chemical potential, with density $\hat{n}$ being the parameter.

In Figure~\ref{fig:entropy_vdW}, the behavior of the entropy in the van der Waals fluid is demonstrated in terms of the appropriate reduced variables:
\begin{equation}
	\rho^*_{\mathrm{vdW}} = \frac{\hat{n}}{3}; \quad T^*_{vdW} = \frac{\hat{\tau}}{4}; \quad \mu^*_{vdW} = \frac{\mu}{4k_{\mathrm B}T_c}
\end{equation}
Note, that in such representation, $v^*_{\mathrm{vdW}} = 2x_c$.

In the van der Waals theory, the concept of entropy per cell formally means the product of the entropy per particle $S^*_{\mathrm{vdW}}$ and density $\rho^*_{\mathrm{vdW}}$, by analogy with the cell model, Eq.~\eqref{eq:S2_via_rho_S1}.




\end{appendices}



\end{document}